\documentclass[12pt]{article}
\usepackage{hyperref}
\usepackage[top=2cm,bottom=3cm,left=2cm,right=2cm]{geometry}
\usepackage{graphicx}
\usepackage{amsmath}
\usepackage{amssymb} 
\usepackage{tikz-cd}

\numberwithin{equation}{section}
\numberwithin{figure}{section}

\def\eq#1{(\ref{eq:#1})}

\def\d{\partial}

\def\H{\mathcal{H}}
\def\Phat{\widehat{\mathcal{P}}}
\def\P{\mathcal{P}}
\def\Ppre{\mathcal{P}_\mathrm{pre}}
\def\deltahat{\widehat{\delta}}
\def\L{\mathcal{L}}

\def\Hloc{\mathcal{H}_\mathrm{loc}}
\def\HP{\mathcal{H}_\mathrm{Peierls}}
\def\Deltacaus{\Delta^\mathrm{causal}}

\begin{document}

\begin{titlepage}
	
	\hfill \today
	\begin{center}
		\vskip 2cm
		
		{\Large \bf {Poisson bracket and $L_\infty$ algebras}
		}
		
		\vskip 0.5cm
		
		\vskip 1.0cm
		{\large {Vin\'{\i}cius Bernardes$^{1}$, Theodore Erler$^{1}$, Atakan Hilmi F{\i}rat$^{2}$, Igor Khavkine$^{3}$ }}
		
		\vskip 0.5cm
		
		{\em  \hskip -.1truecm
			$^{1}$
			CEICO, FZU - Institute of Physics of the Czech Academy of Sciences \\
			No Slovance 2, 182 21, Prague 8, Czech Republic
			\\
			\vskip 0.5cm
			$^{2}$
			Center for Quantum Mathematics and Physics (QMAP),
			Department of Physics \& Astronomy, \\
			University of California, Davis, CA 95616, USA
			\\
			\vskip 0.5cm
			$^{3}$
			Institute of Mathematics of the Czech Academy of Sciences\\
			{\v Z}itn{\' a} 25, 110 00, Prague 1, Czech Republic 
			\\
			\vskip 0.5cm
			\tt \href{mailto:viniciusbernsilva@gmail.com}{viniciusbernsilva@gmail.com},
		\href{mailto:tchovi@gmail.com}{tchovi@gmail.com},\\
		 \href{mailto:ahfirat@ucdavis.edu}{ahfirat@ucdavis.edu}, \href{mailto:khavkine@math.cas.cz}{khavkine@math.cas.cz} \vskip 5pt }
		
		\vskip 2.0cm
		{\bf Abstract}
		
	\end{center}
	\vskip 0.25cm
	\noindent
	\begin{narrower}
		\baselineskip15pt 
		We describe the Poisson bracket of a Lagrangian field theory expressed in the framework of $L_\infty$ algebras. If the symplectic structure on phase space is defined following recent work, we show that the Poisson bracket can be computed with the Peierls formula, confirming the well-known result of the covariant phase space formalism. Of particular interest is the Poisson bracket in nonlocal theories. In $p$-adic string theory we find that a straightforward construction of the Poisson bracket is obstructed by higher derivative instabilities.  We also discuss the inverse relation between the Poisson bracket and symplectic structure in the language of homological algebra, extending some ideas in the mathematical physics literature. 
		
	\end{narrower}
\end{titlepage}

\tableofcontents
\baselineskip15pt

\section{Introduction}

This is part of a series of articles \cite{Bernardes,Bernardes2} describing a new approach to Hamiltonian mechanics based on field theories formulated in terms of $L_\infty$ algebras \cite{Hohm,Jurco}. The goal is to use the covariant phase space formalism \cite{Crnkovic,Crnkovic2,Zuckerman,Lee} and $L_\infty$ data to construct phase space observables without reference to the derivative structure of the Lagrangian. We then can formulate Hamiltonian mechanics in nonlocal field theories where the standard canonical formalism breaks down. Some applications to string field theory \cite{deLacroix,Erler,Erler2,Erbin,Maccaferri,Sen} are discussed in \cite{Bernardes3,Bernardes4,Bernardes5,Bernardes6}. 

The focus here is on the Poisson bracket. We want a construction which inverts the symplectic form 
\begin{equation}
\Omega = \frac{1}{2}\omega(\delta\Phi,[Q_\Phi,\sigma]\delta\Phi),\label{eq:Omega} 
\end{equation}
introduced in \cite{Bernardes}. We show that the relevant inverse is computed using the causal propagator, defining the Poisson bracket as originally described by Peierls \cite{Peierls}. This agrees with the well-known result of the covariant phase space formalism \cite{Forger,Khavkine}, which indirectly implies that the symplectic structure \eq{Omega} should also be standard, at least for finite derivative theories. See also~\cite{Ali}, which explains the origin of \eq{Omega} from the covariant phase space method and establishes a connection to the Barnich-Brandt symplectic form \cite{Barnich}. 

This paper is organized as follows. In section \ref{sec:review}, we summarize the formalism introduced in~\cite{Bernardes,Bernardes2}, and in section \ref{sec:formalism2} we formally define the Poisson bracket. This starts in subsection \ref{subsec:SymObs} with an argument that every physical observable defines a symmetry of the symplectic structure as defined by \eq{Omega}. The symmetry transformation can be constructed via the causal propagator, reproducing the standard Peierls formula. In subsection \ref{subsec:ObsSym} we show that, conversely, every symmetry of the symplectic structure as defined by \eq{Omega} defines a physical observable. The result is an on-shell version of the conserved charge constructed in \cite{Bernardes2}. In subsection \ref{subsec:Poisson} we review the definition and basic properties of the Poisson bracket. In subsection \ref{subsec:symmetry} we discuss the Poisson bracket algebra of conserved charges, giving an explicit formula for the central charge. In section \ref{sec:examples} we discuss two examples. In subsection \ref{subsec:particle} we compute the conserved charges associated to the Galilean symmetry of a nonrelativistic particle, and confirm the presence of a central extension given by the particle's mass. In subsection \ref{subsec:padicPoisson} we investigate the definition of the Poisson bracket in $p$-adic string theory \cite{Brekke} around the unstable vacuum. Higher derivative instabilities lead to an infinite set of poles which prevent a meaningful definition of advanced and retarded propagators. We argue that the theory still has a Poisson bracket if higher derivative modes can be discarded, though it is unclear if it is consistent to do so. Finally in section~\ref{sec:exact} we discuss the inverse relation between the Poisson bracket and symplectic structure from the point of view of homological algebra, a topic of investigation especially in connection to algebraic quantum field theory in curved space. The idea is to realize the causal propagator as the differential of an exact sequence, which in a sense implies that the causal propagator is invertible. The inverse is related to the symplectic form. After discussing this at the level of cohomology in subsection \ref{subsec:iso}, in subsection \ref{subsec:complex} we give an ``off-shell'' generalization which accounts for gauge symmetries. We refer to this as the {\it Peierls complex}. Using the sigmoid we are able to construct an explicit contracting homotopy for the complex, which substantially simplifies and generalizes much previous work on the subject.  

\subsubsection*{Conventions} 

We assume that the products of the $L_\infty$ algebra have odd parity, are graded commutative and carry grade $+1$. All commutators are graded according to whether the object is commuting or anticommuting. 

\section{Review}
\label{sec:review}

In this section we summarize the setup of~\cite{Bernardes,Bernardes2}. The starting point is a vector space $\H$ whose elements can be commuting or anticommuting and carry an integer label called {\it grade}. The vector space comes with a nondegenerate, anticommuting, graded antisymmetric bilinear form at grade $-1$ called the {\it Batalin-Vilkovisky (BV) inner product}, denoted by $\omega$. Inside of $\H$ is the subspace of classical fields $\Phat$:
\begin{equation}\Phi \in \Phat\subseteq \H.\end{equation}
This is the subspace of commuting vectors at grade zero. The basic ingredients of the theory are summarized as follows:  
\begin{description}
\item{\bf Equations of motion:} The dynamics is controlled by an anticommuting element of $\H$ at grade~1 called the {\it Euler-Lagrange state}, $q_\Phi$. The Euler-Lagrange state is a function of the field $\Phi$, and the classical equations of motion are
\begin{equation}q_\Phi = 0.\end{equation}
Solutions to the equations of motion form a nonlinear submanifold of $\Phat$ called {\it pre-phase space} $\Ppre$:
\begin{equation}\Ppre\subseteq\Phat\subseteq\H.\end{equation} 
Variation of $q_\Phi$ defines an anticommuting operator on $\H$ at grade 1 called the {\it kinetic operator},~$Q_\Phi$. If boundary terms can be ignored, the kinetic operator is conserved through the BV inner product
\begin{equation}\omega(Q_\Phi A,B) + (-1)^{|A|}\omega(A,Q_\Phi B) = 0, \end{equation}
where $|A|$ is even or odd when $A$ is commuting or anticommuting. This property is referred to as {\it cyclicity}, and we say that $Q_\Phi$ is {\it cyclic}. We assume that cyclicity can be violated only by boundary terms localized at temporal infinity. Spatial boundary contributions  are arranged to vanish through a suitable definition of the kinetic operator and boundary conditions. The Euler-Lagrange state and kinetic operator together satisfy the Noether identity
\begin{equation}Q_\Phi q_\phi = 0.\end{equation}
The classical action is expressed
\begin{equation}S = -\int_0^1 ds\, \omega\left(\frac{\d\Phi(s)}{\d s},q_{\Phi(s)}\right),\end{equation}
where $\Phi(s)$ depends on a parameter $s\in[0,1]$ with the boundary conditions $\Phi(0)=0$ and $\Phi(1)=\Phi$. The integrand is a total derivative with respect to $s$, so the action only depends on $\Phi(s)$ at $s=1$. If $\deltahat$ is the exterior derivative on $\Phat$, we can show that 
\begin{equation}\deltahat S = \omega(\deltahat\Phi,q_\Phi),\end{equation}
which implies the equations of motion.
\item{\bf Gauge transformations:} We use $\L_\xi$ to denote the Lie derivative along a vector field $\xi$ in $\Phat$ or $\Ppre$. The gauge transformation is defined by the Lie derivative along a vector field $\Lambda$ in $\Phat$ which takes the form 
\begin{equation}\L_\Lambda\Phi = Q_\Phi\lambda_\Phi + \Lambda_\Phi q_\Phi,\end{equation}
where $\lambda_\Phi$ is the {\it gauge parameter} and $\Lambda_\Phi$ is the {\it gauge operator}. Both can be functions of the field, are anticommuting, and carry grade $-1$. The gauge operator is required to be cyclic
\begin{equation}\omega(\Lambda_\Phi A,B) + (-1)^{|A|}\omega(A,\Lambda_\Phi B) = 0,\end{equation}
up to temporal boundary contributions. The gauge operator makes no contribution to the gauge transformation on pre-phase space because the equations of motion are satisfied. The space of solutions modulo gauge transformation is the {\it phase space} $\P$.
\item{\bf Symmetries:}  The Lie derivative along a vector field $\xi$ in $\Phat$ takes the form 
\begin{equation}\L_\xi \Phi = \xi_\Phi, \end{equation}
where $\xi_\Phi$ is a commuting element of $\H$ at grade 0 called the {\it generating parameter}. Variation of $\xi_\Phi$ defines a commuting operator on $\H$ at grade 0 called the {\it generating operator} $\Xi_\Phi$. The generating operator is cyclic
\begin{equation}\omega(\Xi_\Phi A,B) + \omega(A,\Xi_\Phi B) = 0. \end{equation}
Cyclicity can be assumed to hold by definition because it relates the action of $\Xi_\Phi$ at different grades. Both the generating parameter and generating operator can be a functions of the field. The transformation is a symmetry of the action if the following identity holds,
\begin{equation} Q_\Phi \xi_\Phi-\Xi_\Phi q_\Phi=0,\label{eq:symId}\end{equation}
for arbitrary off-shell $\Phi\in\Phat$.
\item{\bf Observables:} An (on-shell) {\it observable} is a function on pre-phase space of the form
\begin{equation}F = -\int_0^1 ds\, \omega\left(\frac{\d\Phi(s)}{\d s},f_{\Phi(s)}\right),\ \ \ \ \Phi\in \Ppre,\label{eq:obs}\end{equation}
where $\Phi(0)=0$ and $\Phi(1)=\Phi\in\Ppre$ is a solution. The object $f_\Phi$ is an anticommuting element of $\H$ at grade 1 called the {\it characteristic state}. The characteristic state is always localized in time to ensure that integration over time does not lead to divergence. Variation of $f_\Phi$ defines the {\it characteristic operator} $F_\Phi$, which is required to be cyclic 
\begin{equation}\omega(F_\Phi A,B) + (-1)^{|A|}\omega(A,F_\Phi B) = 0.\end{equation}
Both the characteristic state and operator can be functions of a solution $\Phi$. If the characteristic state can be extended off-shell in such a way that $F_\Phi$ remains cyclic, the interpolation $\Phi(s)$ can be taken off-shell for intermediate $s$. If $\delta$ is the exterior derivative on pre-phase space, we have an identity analogous to variation of the action
\begin{equation}\delta F = \omega(\delta\Phi,f_\Phi),\label{eq:deltaF}\end{equation}
which implies that the observable only depends on the value of $\Phi(s)$ at $s=1$. 
The observable is gauge invariant if the characteristic state satisfies
\begin{equation}Q_\Phi f_\Phi = 0,\ \ \ \Phi\in\Ppre.\end{equation}
In particular, the observable is a well-defined function on phase space.
\item{\bf Symplectic structure:} Phase space has a symplectic structure \cite{Bernardes}
\begin{equation}
\Omega = \frac{1}{2}\omega(\delta\Phi,[Q_\Phi,\sigma]\delta\Phi),\label{eq:Omega2}
\end{equation}
where $\sigma$ is a commuting operator on $\H$ at grade 0 called the {\it sigmoid} which satisfies boundary conditions in the infinite past and future:
\begin{equation}
\lim_{t\to-\infty}\sigma = 0,\ \ \ \ \lim_{t\to\infty}\sigma = 1.
\end{equation}
Because the exterior derivative on solution space annihilates the Euler-Lagrange state, we must have that $Q_\Phi\delta\Phi = 0$. This almost implies that the symplectic structure vanishes, except that cyclicity of $Q_\Phi$ breaks down because neither $\delta\Phi$ nor $\sigma$ vanishes at temporal infinity. Temporal boundary contributions are critical to ensuring that phase space observables are nontrivial.
\item{\bf Conserved charges:} Given a vector field $\xi$ on $\Phat$ defining a symmetry of the action, the conserved charge can be written \cite{Bernardes2}
\begin{equation}
F_\xi = -\int_0^1 ds\, \omega\left(\frac{\d\Phi(s)}{\d s}, [Q_{\Phi(s)},\sigma]\xi_{\Phi(s)}-[\Xi_{\Phi(s)},\sigma]q_{\Phi(s)} \right).\label{eq:Fv}
\end{equation}
This is an observable defined by the characteristic state
\begin{equation}
f_\Phi = [Q_\Phi,\sigma]\xi_\Phi-[\Xi_\Phi,\sigma]q_\Phi .
\end{equation}
The characteristic operator is cyclic without assuming the equations of motion, so the interpolation $\Phi(s)$ can be taken off-shell. Also useful is the {\it total derivative form} of the conserved charge
\begin{equation}
F_\xi = -\omega(\sigma\xi_{\Phi},q_{\Phi})\big|_\tau,\label{eq:Fvtau}
\end{equation}
where the symbol $|_\tau$ indicates {\it tau regularization}. Tau regularization replaces the field $\Phi$ with $\tau\Phi$ where $\tau$ is an operator that acts as the identity at finite times but vanishes in the infinite past and future. When tau regularization is implemented, contributions from temporal infinity vanish and we can freely assume cyclicity holds.
\end{description}

\section{Formalism}
\label{sec:formalism2}

In this section we describe the construction of the Poisson bracket. Unlike the symplectic structure and conserved charges, the Poisson bracket follows from the definition of the propagator whose existence does not necessarily require a finite derivative Lagrangian. Therefore we do not need a fundamentally new approach to the Poisson bracket. We define it based on the formula of Peierls~\cite{Peierls}. Aside from expressing this in the notation of our formalism, the new result is establishing that the Peierls formula inverts the symplectic structure as defined in~\eq{Omega}.

\subsection{Constructing a symmetry out of an observable}
\label{subsec:SymObs}

\begin{figure}[t]
	\centering
	\includegraphics[scale=1.75]{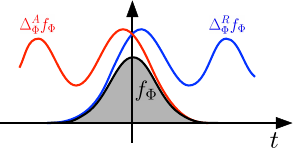}
	\caption{\label{fig:ConsL1} The retarded propagator $\Delta_\Phi^R$ creates a solution which vanishes in the distant past, while the advanced propagator $\Delta_\Phi^A$ creates a solution which vanishes in the distant future, in response to a force $f_\Phi$.} 
\end{figure}

Consider a fluctuation $\varphi$ of a solution $\Phi$ that is being driven by a external force $f_\Phi$. It satisfies the wave equation 
\begin{equation}Q_\Phi \varphi = f_\Phi,\ \ \ \ \Phi\in \Ppre.\end{equation}
The force $f_\Phi$ is an anticommuting state at grade 1. Since the kinetic operator $Q_\Phi$ is nilpotent when $\Phi$ is a solution, $f_\Phi$ must satisfy $Q_\Phi f_\Phi=0$. Further, we will assume that the force is suitably localized in time. Then $f_\Phi$ can be interpreted as the characteristic state of an on-shell observable. There is a general expectation that such a state must be $Q_\Phi$-exact, which means that the fluctuation will be able to respond to the force.\footnote{The expectation follows from the fact that advanced and retarded propagators exist under fairly general conditions with the properties described in section \ref{sec:exact}. An important assumption is that the force is localized in time. Otherwise there can be counterexamples. For example, an electric field cannot consistently adjust to the presence of a point charge on a torus.} This indicates that the kinetic operator has an inverse $\Delta_\Phi$ which satisfies
\begin{equation}Q_\Phi\Delta_\Phi = 1,\label{eq:QDelta}\end{equation} 
when acting on time localized, $Q_\Phi$-invariant states at grade 1. This is the {\it propagator} around the solution $\Phi$. The propagator anticommutes and carries grade $-1$. The solution for the fluctuation field can then be written
\begin{equation}\varphi = \Delta_\Phi f_\Phi.\end{equation}
The propagator is not unique because we can always add to $\varphi$ a solution to the homogeneous equations of motion. In addition, the definition of the propagator requires a choice of gauge. By making use of these freedoms we can construct propagators that generate various kinds of solutions in response to the force. By definition, the {\it retarded propagator} $\Delta_\Phi^R$ creates a solution which starts from $\varphi=0$ in the distant past, responds the force, and subsequently freely evolves. The {\it advanced propagator} $\Delta_\Phi^A$ creates a solution which at first freely evolves but is halted by the force down to $\varphi=0$ in the distant future. See figure~\ref{fig:ConsL1}.  The advanced and retarded propagators are related through the BV inner product
\begin{equation}\omega(\Delta_\Phi^A A,B) = (-1)^{|A|}\omega(A,\Delta_\Phi^R B).\label{eq:DeltaSym}\end{equation}
The {\it causal propagator} is defined as the difference
\begin{equation}\Delta_\Phi^\text{causal} = \Delta_\Phi^R - \Delta_\Phi ^A,\end{equation}
which is annihilated by $Q_\Phi$ on account of \eq{QDelta}
\begin{equation}Q_\Phi\Delta_\Phi^\text{causal} = 0.\end{equation}
The causal propagator is not an inverse for $Q_\Phi$, but conventionally it is still referred to as a propagator.  The causal propagator is cyclic on account of \eq{DeltaSym}:
\begin{equation}
\omega(\Delta_\Phi^\text{causal} A,B) + (-1)^{|A|}\omega(A,\Delta_\Phi^\text{causal}B) = 0.
\end{equation}
We have been imprecise about the definition of advanced and retarded propagators. For local theories the definition is standard and can be made rigorous under suitable conditions. For nonlocal theories there has not been much discussion. We investigate this in the context of $p$-adic string theory in section \ref{subsec:padicPoisson}. 

Noether's theorem states that a symmetry implies the existence of an observable, namely the conserved charge. It is less often emphasized that the reverse is also true. Every observable implies a symmetry. This is the critical fact behind the definition of the Poisson bracket. The precise statement is the following:
\begin{description}
\hypertarget{claim1}{\item{\bf Claim 1:}} Every on-shell observable $F$ can be realized as the conserved charge associated to a Hamiltonian vector field $\xi_F$:
\begin{equation}\iota_{\xi_F}\Omega + \delta F = 0. \end{equation}
The Hamiltonian vector field can be chosen as 
\begin{equation}\L_{\xi_F} \Phi = \Delta_\Phi^\text{causal} f_\Phi,\ \ \ \Phi\in \Ppre, \label{eq:xiF} 
\end{equation}
where $f_\Phi$ is the characteristic state of the observable $F$. Other representatives differ from this by a vector field generating a gauge transformation.
\end{description}
Note that the observable defines a symmetry of the symplectic structure, not necessarily the action. We can interpret the Hamiltonian vector field as the result of ``raising an index'' of the 1-form $\delta F$ using the Poisson structure on phase space. 

Let us give the proof. We start by computing the contraction
\begin{equation}\iota_{\xi_F}\Omega = -\omega(\delta\Phi,[Q_\Phi,\sigma]\Delta_\Phi^\text{causal}f_\Phi).\end{equation}
The characteristic state $f_\Phi$ should be localized to finite time, otherwise the observable $F$ is divergent. However, acting on $f_\Phi$ with a propagator creates a linearized solution in response to a localized force. This solution will not necessarily vanish for large times. Therefore we have to rely on the commutator with the sigmoid to ensure that this expression is localized to finite time. In the next step we will need to open the commutator and use cyclicity, so we implement tau regularization~\cite{Bernardes}. Here it is convenient to insert the regulator $\tau$ only in front of $\delta\Phi$. Opening the commutator and using $Q_\Phi\Delta_\Phi^\text{causal} = 0$ gives
\begin{equation}\iota_{\xi_F}\Omega = -\omega(\tau\delta\Phi,Q_\Phi\sigma\Delta_\Phi^\text{causal}f_\Phi).\end{equation}
We break the causal propagator into advanced and retarded contributions
\begin{equation}
\iota_{\xi_F}\Omega = -\omega(\tau\delta\Phi,Q_\Phi\sigma\Delta_\Phi^R f_\Phi)+\omega(\tau\delta\Phi,Q_\Phi\sigma\Delta_\Phi^A f_\Phi).\label{eq:AR}
\end{equation}
Focus on the advanced term. The advanced propagator creates a solution that vanishes in the infinite future. We can choose the sigmoid to vanish as fast as needed in the past. Therefore the combination $\sigma\Delta_\Phi^A f_\Phi$ is localized in time. We will assume that it is sufficiently well localized to compensate for however fast $\delta\Phi$ might be growing, so that the tau regularization can be dropped. This amounts to a condition on the definition of the advanced propagator. If the tau regulator can be dropped, the contribution from the advanced propagator will vanish because $Q_\Phi\delta\Phi = 0$ after using cyclicity. For the retarded term we add and subtract 1 from the sigmoid to find the expression 
\begin{equation}
\iota_{\xi_F}\Omega = \omega(\tau\delta\Phi,Q_\Phi(1-\sigma)\Delta_\Phi^R f_\Phi)-\omega(\tau\delta\Phi,Q_\Phi\Delta_\Phi^R f_\Phi).
\end{equation}
For the first piece we argue in a similar way as with the advanced term. The retarded propagator creates a solution that vanishes in the infinite past, and we can choose the sigmoid so that $1-\sigma$ vanishes as fast as we want towards the future. Therefore the combination $(1-\sigma)\Delta_\Phi^R f_\Phi$ is localized in time. We assume that it is sufficiently localized that the tau regularization to be dropped, which places a condition on the definition of the retarded propagator. Then we find zero after using cyclicity of $Q_\Phi$. In the remaining term we note
\begin{equation}Q_\Phi\Delta^R_\Phi f_\Phi = f_\Phi,\end{equation}
which implies 
\begin{equation}\iota_{\xi_F}\Omega= -\omega(\delta\Phi, f_\Phi) = -\delta F.\end{equation}
This completes the proof.

Let us explain why the symplectic form is nondegenerate. Nondegeneracy means that the contraction $\iota_\xi\Omega$ completely determines the vector field $\xi$ on phase space. To see why this is true, suppose we contract $\iota_\xi\Omega$ with the Peierls vector field $\xi_F$ of an on-shell observable $F$. By claim 1, we know that 
\begin{equation}\iota_{\xi_F}\iota_\xi \Omega = \L_\xi F.\end{equation}
Since $F$ can be an arbitrary function on phase space, knowing the Lie derivative completely determines the vector field $\xi$ on phase space. In \cite{Vitagliano} it has been noted that under some cohomological conditions, without requiring the existence of the advanced and retarded propagators, the kernel of $\Omega$ can already be identified with gauge transformations. 

In the framework of $L_\infty$ algebras we often work perturbatively, and it is of interest to understand how the propagator $\Delta_\Phi$ around a solution $\Phi$ can be related back to the propagator $\Delta_0$ around the vacuum $\Phi=0$. Assuming that the propagator acts on $Q_\Phi$-invariant states at grade 1, there is a nice formula
\begin{equation}\Delta_\Phi = \Delta_0 \frac{1}{Q_\Phi\Delta_0}.\end{equation}
It is easy to see that this formally satisfies the critical relation \eq{QDelta}, but it is necessary to define the inverse of $Q_\Phi\Delta_0$. Around $\Phi=0$ we assume that $\Delta_0$ already satisfies \eq{QDelta},
\begin{equation}Q\Delta_0 = 1,\end{equation}
when acting on time-localized, $Q$-invariant states at grade 1. Therefore
\begin{equation}Q_\Phi\Delta_0= 1+(Q_\Phi-Q)\Delta_0 \ \ \ \ \text{(grade 1)}.\end{equation}
The second term starts at linear order in $\Phi$ and can be treated as a small correction to the identity operator. Therefore  $Q_\Phi\Delta_0$ can be inverted using a geometric series. 

\subsection{Constructing an observable out of a symmetry}
\label{subsec:ObsSym}

The inverse statement to \hyperlink{claim1}{claim 1} concerns the construction of observables from symmetries. This is mainly the result of Noether's theorem, and was discussed in \cite{Bernardes2}. However Noether's theorem applies to symmetries of the action. The symplectic structure can have symmetries that are not realized in the action, such as electromagnetic duality in Maxwell theory. To account for these we need a slight generalization of the result of \cite{Bernardes2}: 
\begin{description}
\hypertarget{claim2}{\item{\bf Claim 2:}} Every symmetry of the symplectic structure can be realized (locally) as a Hamiltonian vector field $\xi$ associated to a conserved charge $F_\xi$
\begin{equation}\iota_\xi\Omega + \delta F_\xi = 0.\label{eq:ham}\end{equation}
The conserved charge takes the form
\begin{equation}
F_\xi = -\int_0^1 ds\,\omega\left(\frac{\d\Phi(s)}{\d s},[Q_{\Phi(s)},\sigma]\xi_{\Phi(s)}\right),\ \ \ \ \Phi(s)\in \Ppre, \label{eq:Fxionshell}
\end{equation}
where $\xi_\Phi$ is the generating parameter of the symmetry and the interpolation $\Phi(s)$ must be a solution for all $s$.
\end{description}
The utility of this result is limited by the fact that it requires finding a path through the space of solutions to $\Phi=0$. This is one reason why conserved charges are usually derived from symmetries of the action. 

Let us give a proof. To solve for the conserved charge we note 
\begin{equation}
\iota_\xi\Omega = -\omega(\delta\Phi,f_\Phi),\label{eq:323}
\end{equation}
where 
\begin{equation}f_\Phi = [Q_\Phi,\sigma]\xi_\Phi.\label{eq:Qsigmaxi}\end{equation}
Because $\xi$ must be a symmetry,
\begin{equation}\L_\xi\Omega = 0,\end{equation}
the contraction \eq{323} must be $\delta$-closed. To check whether it is, we introduce the operator $F_\Phi$ through
\begin{equation}\delta f_\Phi = -F_\Phi\delta\Phi.\label{eq:325}\end{equation}
Because $\Phi$ is a solution, $F_\Phi$ is defined on $Q_\Phi$-invariant states at grade 0. Computing the exterior derivative of \eq{323}, $\xi$ will be a symmetry of the symplectic structure if
\begin{equation}\L_\xi\Omega = \omega(\delta\Phi,F_\Phi\delta\Phi) = 0.\end{equation}
Expanding $\delta\Phi$ in basis 1-forms and using graded antisymmetry of $\omega$, it quickly follows that $F_\Phi$ is cyclic. Therefore if $\xi$ is a symmetry of the symplectic structure, $f_\Phi$ in \eq{Qsigmaxi} and
$F_\Phi$ in \eq{325} have the necessary properties to define the characteristic state and characteristic operator of an observable. However, we should be aware of the fact that in this case the characteristic state and operator are {\it a priori} only defined on-shell. It follows from \eq{323} that the conserved charge is given by \eq{obs} where $\Phi(s)$ follows a path through the space of solutions. Then \eq{Fxionshell} follows.

In this way we find a kind of isomorphism between symmetries and observables. Given the characteristic state of an observable $f_\Phi$, \hyperlink{claim1}{claim 1} states that we can derive the generating parameter of a symmetry $\xi_\Phi$ through
\begin{equation}\xi_\Phi = \Deltacaus_\Phi f_\Phi.\end{equation}
Conversely, given the generating parameter of a symmetry $\xi_\Phi$, \hyperlink{claim2}{claim 2} implies that we can derive the characteristic state of an observable $f_\Phi$ through 
\begin{equation}f_\Phi = [Q_\Phi,\sigma]\xi_\Phi.\end{equation}
In a certain sense $[Q_\Phi,\sigma]$ appears to be an inverse of the causal propagator $\Deltacaus_\Phi$. Making this observation precise requires some tools in homological algebra, which we discuss in section \ref{sec:exact}.

\subsection{Poisson bracket}
\label{subsec:Poisson}

The Poisson bracket of on-shell observables $F,G$ is defined by
\begin{equation}[F,G] = \L_{\xi_G} F,\label{eq:Poisson}\end{equation}
where $\xi_G$ is a Hamiltonian vector field whose conserved charge is $G$. We may construct $\xi_G$ through the causal propagator as \eq{xiF}, but this is not essential because any other Hamiltonian vector field would differ from this by a gauge transformation which leaves $F$ invariant. Using \eq{xiF} and the form of the variation of $F$, it quickly follows that the Poisson bracket is given by
\begin{equation}
[F,G] = \omega(f_\Phi, \Delta_\Phi^\mathrm{causal} g_\Phi),
\end{equation}
where $f_\Phi$ is the characteristic state associated to $F$ and $g_\Phi$ is the characteristic state associated to~$G$. This is how we write the standard Peierls formula \cite{Peierls} in our formalism. Note that the construction of the Poisson bracket depends nontrivially on $\Phi$, i.e. on where we are in phase space. This is different from the standard presentation in the canonical formalism, where the Poisson bracket takes the same form regardless of $p$ and $q$. This is a special consequence of the fact that $p$s and $q$s form Darboux coordinates on phase space. 

The Poisson bracket has a number of properties:
\begin{description}
\item{\bf Antisymmetry:}
\begin{equation} [F,G] = -[G,F].\end{equation}
\item{\bf Linearity:} $[F,G]$ is linear in $F$ and $G$. 
\item{\bf Gauge invariance:} The Poisson bracket is gauge invariant,
\begin{equation}\L_\Lambda[F,G] = 0,\end{equation}
where $\Lambda$ is a vector field on $\Ppre$ generating a gauge transformation. This means that the Poisson bracket of on-shell observables is itself an on-shell observable.
\item{\bf Leibniz rule:} The Poisson bracket acts as a derivation of the algebra of on-shell observables:
\begin{equation}[F,GH] = [F,G]H + G[F,H].\end{equation}
\item{\bf Jacobi identity:}  
\begin{equation}[F,[G,H]]+[G,[H,F]]+[H,[F,G]]=0.\end{equation}
This implies that on-shell observables form a Lie algebra whose bracket is the Poisson bracket. 
\end{description}
These properties follow from Cartan calculus and do not rely on anything unique to our development. However it is useful to demonstrate gauge invariance since the question is specific to the covariant phase space formalism. Using \eq{Poisson} we can compute
\begin{equation}
\L_\Lambda [F,G] = \L_\Lambda\L_{\xi_G} F = [\L_\Lambda,\L_{\xi_G}] F = \L_{[\Lambda,\xi_G]} F.
\end{equation}
To understand the nature of the Lie bracket vector field $[\Lambda,\xi_G]$ we contract with the symplectic form
\begin{equation}
\iota_{[\Lambda,\xi_G]}\Omega = [\iota_\Lambda,\L_{\xi_G}] \Omega = \iota_\Lambda\L_{\xi_G}\Omega-\L_{\xi_G}\iota_\Lambda\Omega.
\end{equation}
The first term vanishes because $\xi_G$ generates a symmetry of the symplectic structure, and the second term vanishes because $\Lambda$ generates a gauge transformation. This means that the Lie bracket vector field $[\Lambda,\xi_G]$ also generates a gauge transformation which leaves the observable $F$ invariant. Therefore the Poisson bracket is gauge invariant.

\subsection{Symmetry algebras}
\label{subsec:symmetry}

An important application of the Poisson bracket is the expression of symmetry algebras. This is a consequence of the following:
\begin{description}
\item{\bf Claim 3:} Let $\xi$ and $\gamma$ be vector fields on $\Phat$ which generate symmetries of the action, and let $F_{\xi},F_{\gamma}$ be the respective conserved charges. Then
\begin{equation}
[F_{\xi},F_{\gamma}] + F_{[\xi,\gamma]} = C_{\xi,\gamma},\label{eq:PBalgebra}
\end{equation}
where $F_{[\xi,\gamma]}$ is the conserved charge corresponding to the Lie bracket of vector fields and $C_{\xi,\gamma}$ is the {\it central charge}. It is locally a constant on phase space given by
\begin{equation}
C_{\xi,\gamma} = \omega\big(\xi_\Phi,[Q_\Phi,\sigma]\gamma_\Phi\big)\Big|_{\Phi=0},\label{eq:Cxigamma}
\end{equation}
where $\xi_\Phi$ and $\gamma_\Phi$ are the generating parameters of the symmetries $\xi$ and $\gamma$.
\end{description}
The central charge is an antisymmetric bilinear form on tangent vectors to $\Ppre$, 
\begin{equation}C_{\xi,\gamma} = -C_{\gamma,\xi},\end{equation}
which satisfies 
\begin{equation}
C_{\xi,[\gamma,\theta]}+C_{\gamma,[\theta,\xi]}+C_{\theta,[\xi,\gamma]}=0.
\end{equation}
This means that the central charge is a 2-cocycle in the language of Lie algebra cohomology. In our definition \eq{Fv}, the conserved charge always starts at linear order in $\Phi$. One can of course add a constant to the charge and it will still be conserved. By doing this it may be possible to eliminate the central charge from the Poisson bracket algebra. This will happen if the central charge depends on $\xi$ and $\gamma$ only as a linear function of the Lie bracket $[\xi,\gamma]$. In the language of Lie algebra cohomology, the central charge would then be a 2-coboundary. If it depends on $\xi$ and $\gamma$ in a more general way, then the central charge cannot be removed. Then the symmetry algebra has a nontrivial central extension. 

The above result follows from Cartan calculus and the symplectic structure \eq{Omega}. However, it is instructive to demonstrate it by directly evaluating the Poisson bracket of the conserved charges as given  by~\eq{Fv}. From the definition \eq{Poisson} 
\begin{align}
[F_\xi,F_\gamma] & = \L_\gamma F_\xi \nonumber\\
& = -\omega\big(\L_\gamma\Phi,[Q_\Phi,\sigma]\xi_\Phi-[\Xi_\Phi,\sigma]q_\Phi\big).
\end{align}
In the second step we used \eq{deltaF},  and in the second entry of $\omega$ appears the characteristic state of the conserved charge $F_\xi$. Since we are on-shell the Euler-Lagrange state is zero, and substituting the generating parameter of $\gamma$ gives
\begin{equation}
[F_\xi,F_\gamma] = \omega(\xi_\Phi,[Q_\Phi,\sigma]\gamma_\Phi).
\end{equation}
Note that the Poisson bracket in this case does not involve the causal propagator. This is because the conserved charges come from symmetries that we already know, and it is not necessary to construct them via \eq{xiF}. We extract the field-independent part of the Poisson bracket by writing
\begin{equation}
[F_\xi,F_\gamma] = \omega(\xi_\Phi,[Q_\Phi,\sigma]\gamma_\Phi)\Big|_{\Phi=0} + \omega(\xi_\Phi,[Q_\Phi,\sigma]\gamma_\Phi)\Big|_{\Phi=0}^\Phi,
\end{equation}
where the second term indicates the difference between $\Phi$ and $\Phi=0$. The first term is the central charge. The second term is at least linear on $\Phi$, so we are able to open the commutator and invoke cyclicity if we implement tau regularization $\Phi\to\tau\Phi$. The result is
\begin{equation}
[F_\xi,F_\gamma] = C_{\xi,\gamma}-\Big[\omega(Q_\Phi\xi_\Phi,\sigma \gamma_\Phi)\big|_\tau+\omega(\xi_\Phi,\sigma Q_\Phi\gamma_\Phi)\big|_\tau\Big] \Big|_{\Phi=0}^\Phi.
\end{equation}
Because $\xi$ and $\gamma$ are symmetries of the action we have
\begin{equation}
Q_\Phi \xi_\Phi-\Xi_\Phi q_\Phi = 0,\ \ \ \ Q_\Phi \gamma_\Phi - \Gamma_\Phi q_\Phi = 0,\ \ \ \ \Phi\in\Phat,
\end{equation}
where $\Gamma_\Phi$ is the generating operator corresponding to $\gamma$. Substituting, we obtain
\begin{equation}
[F_\xi,F_\gamma] = C_{\xi,\gamma}-\Big[\omega(\Xi_\Phi q_\Phi,\sigma \gamma_\Phi)\big|_\tau+\omega(\xi_\Phi,\sigma \Gamma_\Phi q_\Phi)\big|_\tau\Big] \Big|_{\Phi=0}^\Phi.
\end{equation}
Since we assume that $\Phi=0$ satisfies the equations of motion, we can drop the $\Phi=0$ term even when there is a $\tau$ regulator. We use cyclicity of the generating operators to write 
\begin{align}
[F_\xi,F_\gamma] = C_{\xi,\gamma}+ & \omega(q_\Phi,[\Xi_\Phi,\sigma]\gamma_\Phi)\big|_\tau - \omega(q_\Phi,[\Gamma_\Phi,\sigma]\xi_\Phi)\big|_\tau\nonumber\\
& +\omega(q_\Phi,\sigma \Xi_\Phi\gamma_\Phi)\big|_\tau - \omega(q_\Phi,\sigma \Gamma_\Phi\xi_\Phi)\big|_\tau.
\end{align}
The commutator terms are localized to finite time and drop out by the equations of motion. This leaves
\begin{equation}
[F_\xi,F_\gamma] = C_{\xi,\gamma} +\omega\big(q_\Phi,\sigma (\Xi_\Phi\gamma_\Phi- \Gamma_\Phi\xi_\Phi)\big)\Big|_\tau.
\end{equation}
The next step is to identify the generating state of the commutator vector field
\begin{equation}
\L_{[\xi,\gamma]}\Phi = [\L_\xi,\L_\gamma]\Phi = \L_\xi \gamma_\Phi -\L_\gamma\xi_\Phi=\Gamma_\Phi\xi_\Phi -\Xi_\Phi\gamma_\Phi,\ \ \ \ \Phi\in\Phat.
\end{equation}
With this we recognize the second term as the total derivative form of the conserved charge $F_{[\xi,\gamma]}$, written in \eq{Fvtau}. The result follows.

\section{Examples}
\label{sec:examples}

In this section we present some calculations involving the Poisson bracket. The derivation of the causal propagator in local field theories is in principle standard, and the Peierls formula is discussed extensively for example in work of DeWitt \cite{DeWitt}. Therefore we will not work out many local field theory examples, limiting ourselves to the point particle. See also \cite{Harlow}. In $p$-adic string theory the construction of the Poisson bracket has not been discussed and raises interesting questions. We apply the formulas using DeWitt notation following appendix C of \cite{Bernardes2}. 

\subsection{Galilean algebra for particle}
\label{subsec:particle}

Consider a nonrelativistic particle with the action
\begin{equation}S = \frac{m}{2}\int dt\, \dot{X}^i\dot{X}^i,\end{equation}
where $X^i$ is the position of the particle as a function of time $t$, $i$ is a spatial coordinate index,  the dot indicates the time derivative, and $m$ is the mass of the particle. Integration over $t$ is carried out from minus infinity to infinity. Following the prescription of \cite{Bernardes2}, we expand the action in powers of a fluctuation and read off the Euler-Lagrange state and kinetic operator
\begin{equation}
q^i(t) = m\ddot{X}^i(t),\ \ \ \ Q^{ij}(t,t') = m\delta^{ij}\frac{d^2}{dt^2}\delta(t-t').\label{eq:RQpart}
\end{equation}
The particle has Galilean symmetry. We can derive the conserved charges following \cite{Bernardes2} or by standard means to arrive at  
\begin{subequations}
\begin{align}
H & = \frac{m}{2}\int dt\, \dot{\sigma} \dot{X}^i\dot{X}^i,\\
P^i & = m\int dt\,\dot{\sigma}\dot{X}^i,\\
L^{ij} & = m\int dt\, \dot{\sigma} X^{[i}\dot{X}^{j]},\\
B^i & = m\int dt\,\dot{\sigma}\big(t \dot{X}^i-X^i\big),
\end{align}
\end{subequations}
where $H$ is the Hamiltonian, $P^i$ is the momentum, $L^{ij}$ is the angular momentum, and $B^i$ is the Galilean boost charge. We assume that the sigmoid acts through multiplication by $\sigma(t)$. If the sigmoid is a unit step function, it is easy to recognize the grade school formulas for energy, momentum, and angular momentum. Note that the boost charge depends explicitly on time. This means that it will not be invariant under time translations of the solution. It is important to understand that a conserved charge is not necessarily time independent. For this it must commute with the Hamiltonian. Conservation means independence from the choice of sigmoid. The Galilean boost charge is conserved in this sense.  

\begin{figure}[t]
	\centering
	\includegraphics[scale=1.6]{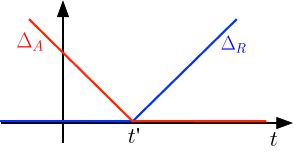}
	\caption{\label{fig:ConsL2} Advanced and retarded propagators for the nonrelativistic particle.} 
\end{figure}

The retarded propagator should satisfy
\begin{equation}\int dt_2\,Q^{ij}(t_1,t_2)\Delta_R^{jk}(t_2,t_3) = \delta^{ik}\delta(t_1-t_3).\end{equation}
This should hold when $t_3$ is integrated against the characteristic state of an observable, but because the particle has no gauge symmetry every sufficiently localized function of $t_3$ represents an observable. Substituting $Q^{ij}(t_1,t_2)$ this becomes
\begin{equation}
m\frac{d^2}{dt^2}\Delta_R^{ij}(t,t') = \delta^{ij}\delta(t-t').
\end{equation}
Integrating twice while recalling that the retarded propagator should vanish in the past gives
\begin{equation}
\Delta^{ij}_R(t,t') = \frac{\delta^{ij}}{m}(t-t')\theta(t-t'),
\end{equation}
where $\theta$ is the Heaviside step function. In a similar way the advanced propagator is found to be
\begin{equation}\Delta^{ij}_A(t,t') = -\frac{\delta^{ij}}{m}(t-t')\theta(t'-t).\end{equation}
See figure \ref{fig:ConsL2}. Note that the retarded propagator not only vanishes in the infinite past, but is strictly zero before it receives an impulse at $t=t'$. This is generally expected for a local theory, because the field cannot respond to an impulse before it occurs. The advanced and retarded propagators are related by exchange 
\begin{equation}\Delta^{ij}_R(t,t') = \Delta^{ji}_A(t',t),\end{equation}
in agreement with \eq{DeltaSym}. The causal propagator is 
\begin{equation}\Delta^{ij}_\text{causal}(t,t') = \frac{\delta^{ij}}{m}(t-t'),\end{equation}
which means that the Poisson bracket can be computed as
\begin{equation}
[F,G] = -\frac{1}{m}\int dt_1dt_2 (t_1-t_2)f^i(t_1)g^i(t_2),
\end{equation}
where $f^i$ and $g^i$ are the characteristic states of the observables $F$ and $G$.

To see that this formula makes sense, we confirm that position and momentum have the canonical Poisson bracket
\begin{equation}[X^i(0),P^j] = \delta^{ij},\end{equation}
where $X^i(0)$ is the position of the particle at $t=0$. The position is an observable defined by the characteristic state
\begin{equation}f_{X^j(0)}^i(t) = -\delta^{ij}\delta(t),\end{equation}
while the momentum is defined by the characteristic state 
\begin{equation}f_{P^j}^i(t) = m\delta^{ij}\ddot{\sigma}.\end{equation}
The Poisson bracket is
\begin{align}
[X^i(0),P^j] & = -\frac{1}{m}\int dt_1dt_2 (t_1-t_2)f_{X^i(0)}^k(t_1)f_{P^j}^k(t_2)\nonumber\\
& =  \frac{1}{m}\int dt_1dt_2 (t_1-t_2) \delta^{ik}\delta(t_1) \times m\delta^{jk} \ddot{\sigma}(t_2) \nonumber\\
& = -\delta^{ij}\int dt\, t \ddot{\sigma}.
\end{align}
It is useful to pull out a total derivative in the integrand so that we write
\begin{equation}
[X^i(0),P^j] = \delta^{ij}\int dt\, \left[\dot{\sigma} - \frac{d}{dt}(t\dot{\sigma})\right].
\end{equation}
Using
\begin{equation}\int dt\, \dot{\sigma} = \sigma(\infty)-\sigma(-\infty) = 1,\label{eq:intsigmadot}\end{equation}
the first term in the integrand gives 1, reproducing the expected result. The second term does not contribute because $\dot{\sigma}$ must vanish faster than $1/t$ for large times. This follows because the sigmoid itself does not grow logarithmically. 

If we compute the Poisson brackets of the conserved charges we should obtain the Galilean symmetry algebra. We will not calculate all the Poisson brackets but a noteworthy example concerns the translation and boost generators. The characteristic state of the boost generator is
\begin{equation}f_{B^j}^i(t) = m\delta^{ij}\frac{d^2}{dt^2}(t\sigma),\end{equation}
from which we can calculate the Poisson bracket
\begin{align}
[P^i,B^j] & = -\frac{1}{m}\int dt_1 dt_2 (t_1-t_2)f_{P^i}^k(t_1)f_{B^j}^k(t_2)\nonumber\\
& = -\frac{1}{m}\int dt_1 dt_2(t_1-t_2) m\delta^{ik}\ddot{\sigma}(t_1)\times m\delta^{jk}\frac{d^2}{dt_2^2}\big(t_2\sigma(t_2)\big)\nonumber\\
& = -m\delta^{ij} \int dt_1 dt_2(t_1-t_2)\ddot{\sigma}(t_1)\big(t_2\ddot{\sigma}(t_2)+2\dot{\sigma}(t_2)\big).
\end{align}
Integrating by parts in $t_1$ we eliminate the $t_1-t_2$ factor from the causal propagator. There is no boundary term again because $\dot{\sigma}$ vanishes faster than $1/t$. The integrand then factorizes to 
\begin{equation}
[P^i,B^j] = m\delta^{ij}\left(\int dt_1\dot{\sigma}(t_1)\right)\left(\int dt_2\big(t_2\ddot{\sigma}(t_2)+2\dot{\sigma}(t_2)\big)\right).
\end{equation}
The $t_1$ integral evaluates to $1$. In the $t_2$ integral we integrate by parts on the $\ddot{\sigma}$ term to find
\begin{equation}
[P^i,B^j] = m\delta^{ij}\int dt_2\big(-\dot{\sigma}(t_2)+2\dot{\sigma}(t_2)\big).
\end{equation}
Canceling finally gives 
\begin{equation}
[P^i,B^j] = m\delta^{ij}.
\end{equation}
The mass of the particle appears as a central charge. The central extension of Galilean algebra is known as the Bargmann algebra.

\subsection{Poisson bracket in $p$-adic string theory}
\label{subsec:padicPoisson}

\begin{figure}[t]
	\centering
	\includegraphics[scale=.35]{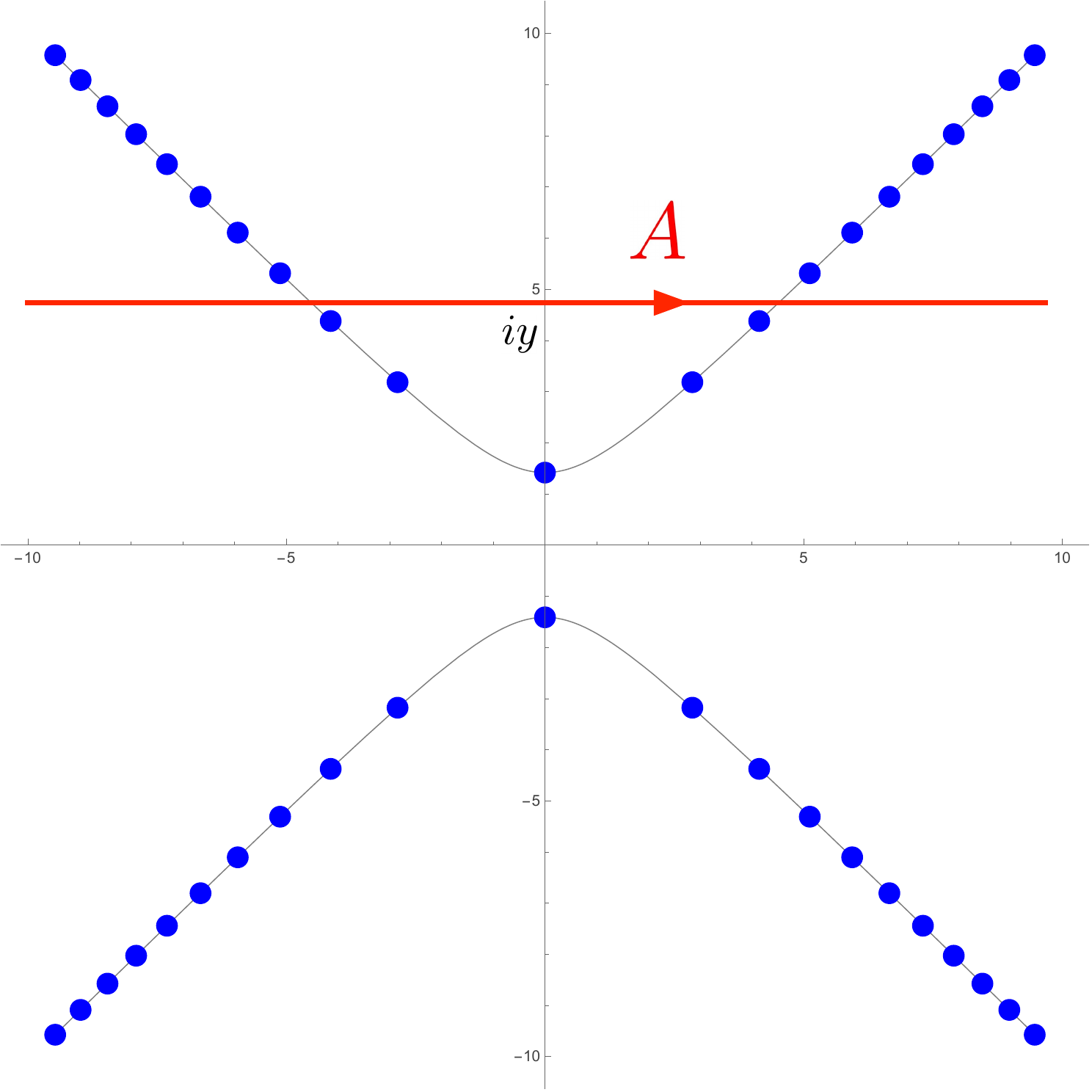}
	\caption{\label{fig:ConsL4} The blue dots represent the poles of the propagator in $p$-adic string theory expanded around the unstable vacuum $\phi=1$. There is no choice of contour for the advanced propagator that goes above all of the poles.} 
\end{figure}

We consider the Poisson bracket in $p$-adic string theory \cite{Brekke} around the unstable vacuum $\phi=1$. Restricting to one time dimension $t$, the kinetic operator is
\begin{equation}
Q(t,t')= \frac{1}{g^2}\Big(p^{\frac{1}{2}\frac{d^2}{dt^2}}-p\Big)\delta(t-t').
\end{equation}
The advanced propagator should be constructed to satisfy
\begin{equation}
\int dt_2 \,Q(t_1,t_2)\Delta_A(t_2,t_3) = \delta(t_1-t_3),
\end{equation}
subject to the condition that it vanishes fast enough in the future. This gives
\begin{equation}
\Delta_A(t,t') = \int_A\frac{dE}{2\pi}\frac{e^{iE(t-t')}}{K(E)},
\end{equation} 
where 
\begin{equation}
K(E) = \frac{1}{g^2}\big(p^{-E^2/2}-p\big)
\end{equation}
and $A$ would normally be chosen as the contour which follows parallel to the real axis but above all poles of $1/K(E)$. With this choice, we can close the contour in the upper half plane when $t>t'$ without encircling any poles, and the advanced propagator will be strictly zero when $t$ is in the future of~$t'$. The problem is that $1/K(E)$ has an infinite number of poles with unbounded imaginary part. This is shown in figure \ref{fig:ConsL4}. There is no contour $A$ which follows parallel to the real axis that goes above all of them. Instead one can think about a contour which approaches infinity at a $45^\circ$ angle above all of the poles. Computing
\begin{equation}
\int dt' Q(t,t')\Delta_A(t',t'') = \int_A\frac{dE}{2\pi} e^{iE(t-t'')}
\end{equation}
is supposed to give a delta function $\delta(t-t'')$. But unfortunately the contour approaches infinity in the wrong direction, and the integral appears to diverge when $t<t''$. To get a delta function the endpoints of the contour $A$ need to be pinned to the real axis at infinity.

The poles are located on two branches of a hyperbola defined by the relation
\begin{equation}(E_I)^2 = -2\left(1+\frac{2\pi i n}{\ln p}\right),\end{equation}
where $I$ labels the poles and $n$ is a corresponding integer. The two poles with $n=0$ represent a tachyon with mass-squared
\begin{equation}m^2 = -2.\label{eq:p-tachyon}\end{equation}
This can be interpreted as a conventional tachyonic instability of the D-brane in $p$-adic string theory. The remaining poles represent unconventional particle states with complex mass-squared. These indicate higher derivative instabilities and if taken seriously probably indicate that $p$-adic string theory is unphysical.  

While there is no acceptable contour that passes above all of the poles, we can think about a contour that passes above some of them but not others. This leads to a propagator
\begin{align} 
\Delta_A(t,t') & = \theta(t-t')\left(\sum_{E_I\text{ above } A}\frac{ie^{i E_I(t-t')}}{\d K(E_I)}  \right)-\theta(t'-t)\left(\sum_{E_I\text{ below } A}\frac{ie^{i E_I(t-t')}}{\d K(E_I)}\right).\label{eq:Cyprop}
\end{align}
The large $t$ behavior is 
\begin{equation}\Delta_A(t,t') \sim \frac{i}{\d K(E_\mathrm{min})} e^{i E_\mathrm{min}(t-t')},\label{eq:larget}\end{equation}
where $E_\mathrm{min}$ is the pole with smallest imaginary part above the contour $A$. Therefore the advanced propagator will vanish in the infinite future as needed provided that $E_\mathrm{min}$ has positive imaginary part. This suggests we can take the contour $A$ to be on the real axis or anywhere above. However this cannot be quite right, because if $A$ is on the real axis then the advanced propagator will also vanish in the infinite past. Therefore the propagator appears to be both advanced and retarded, and the Poisson bracket will be strictly zero.  

Apparently it is not enough to require that the advanced propagator vanishes towards the future. To understand what is needed we must revisit the argument of subsection \ref{subsec:ObsSym}, which ensures that we construct a symmetry out of an observable. The crucial point is explained below \eq{AR}, where it is assumed that the advanced propagator vanishes fast enough in the future that the expression 
\begin{equation}\omega(\tau\delta\Phi,Q_\Phi\sigma\Delta^A_\Phi f_\Phi)\end{equation}
vanishes. In the present context the expression takes the form
\begin{equation}
\omega(\tau\delta\Phi,Q_\Phi\sigma\Delta^A_\Phi f_\Phi) = -\int dt_1 dt_2 dt_3\, \tau(t_1)\delta\phi(t_1) Q(t_1,t_2)\sigma(t_2)\Delta_A(t_2,t_3)f(t_3),\label{eq:an_exp}
\end{equation}
where the sigmoid and tau regularization operate through multiplication by functions $\sigma(t),\tau(t)$ and $f(t)$ is the characteristic state of an observable. Because the kinetic operator annihilates $\delta\phi(t)$, this expression only gets contribution from the distant past and future where the tau regulator transitions between 0 and 1. Let us say that the transition occurs at a time $t_1\approx \pm T$ for some very large $T$. Let us further assume that solutions can grow at fastest as $e^{\mu t}$ in the distant future for some $\mu\geq 0$, and the sigmoid vanishes fast enough in the past that we can drop contributions from $-T$. Finally, let us assume that the characteristic state of the observable is a delta function at $t=0$. Then \eq{an_exp} will be proportional to 
\begin{equation}
\omega(\tau\delta\Phi,Q_\Phi\sigma\Delta^A_\Phi f_\Phi) \, \propto\  e^{\mu T}\Delta_A(T,0).
\end{equation}
This will be zero if the advanced propagator vanishes towards the future faster than any admissible solution grows. From \eq{larget} we see that the pole with smallest imaginary part above the advanced contour must satisfy 
\begin{equation}\mathrm{Im}(E_\text{min})>\mu. \end{equation}
By a similar argument, the retarded propagator must be chosen so that the pole with least negative imaginary part below the retarded contour must have more negative imaginary part than $-\mu$. 

Unfortunately the poles of the propagator tell us how fast solutions can grow, and from this we see that $\mu$ is unbounded. Therefore there appears to be no definition of the Poisson bracket that accounts for all solutions. If we believe that the theory must have a Poisson bracket, we can only keep solutions whose poles are below the advanced contour and above the retarded contour. One reasonable suggestion is to keep the tachyon \eq{p-tachyon} while discarding the infinite sequence of higher derivative modes. Such a truncation has already been suggested in~\cite{Barnaby}. Indeed, the nonlocal action of $p$-adic string theory is designed specifically to reproduce the $p$-adic Koba-Nielsen amplitudes involving the tachyon. Amplitudes involving the higher derivative modes are of questionable significance. Still, it is not clear that these modes can be consistently ignored. In perturbation theory the nonlinear equations of motion can be made second order either by integrating out the higher derivatives \cite{Eliezer} or by removing them through field redefinition~\cite{Erbin3}. Perhaps such a prescription is still sensible nonperturbatively.

It is interesting to mention that a very similar spectrum of poles appears in open string field theory expanded around the tachyon vacuum at level zero \cite{Kostelecky}. In this case all poles have complex mass-squared and appear to be associated to higher derivative instabilities. If these modes are discarded, the phase space at the tachyon vacuum becomes zero dimensional in accordance with Sen's conjecture \cite{Sen2}. However, this is only the action at level zero. It is unclear whether the poles must be removed by hand once the full cohomology problem is analyzed at higher levels. The existence of a contracting homotopy associated to Schnabl's analytic solution \cite{Schnabl,Ellwood} suggests the complete absence of cohomology, but other computations appear to contradict this \cite{Imbimbo,Kishimoto,Inatomi}. The question is important for understanding whether a Poisson bracket exists for nonperturbative solutions in string field theory.

\section{Inverse of the causal propagator}
\label{sec:exact}

Understanding how the Poisson bracket inverts the symplectic structure is related to the problem of defining an inverse for the causal propagator. We have seen that, roughly speaking, the inverse is~$[Q_\Phi,\sigma]$. However, the causal propagator is not strictly invertible because it is annihilated by~$Q_\Phi$. One can think about defining an inverse after taking a suitable quotient. Here it starts to be natural to think about the problem from the point of view of homological algebra. There is a long history to this line of thinking, especially in literature on algebraic quantum field theory in curved space. The objective is to invert the causal propagator by realizing it as the differential of an exact sequence. The standard construction is described in ~\cite{Benini2} (equation (3.8) or (8) in arXiv version), with precursors in \cite{WaldQFT} (lemma 3.2.1), \cite{Bar} (theorem 3.4.7), and \cite{Khavkine} (proposition 2.1). In this section we articulate a generalization which accounts for gauge symmetries. With the help of the sigmoid, we are able to construct an explicit contracting homotopy for the sequence, which is directly related to the symplectic structure. We refer to this as the {\it Peierls complex}, and its construction unifies and simplifies much previous work on the topic.

To define the complex we need three subspaces of $\H$:
\begin{align}
 \text{Retarded subspace},\ \H_R \ \ \ \ &\rightarrow\ \ \ \text{Space of fields which vanish quickly in the past,}\nonumber\\
\text{Advanced subspace},\ \H_A \ \ \ &\rightarrow \ \ \ \text{Space of fields which vanish quickly in the future,}\nonumber\\
 \text{Localized subspace},\ \Hloc \ \ \ &\rightarrow \ \ \ \text{Space of fields which vanish quickly in both past and future.}\nonumber
\end{align}
The full vector space can be regarded as the union 
\begin{equation}\H = \H_R+\H_A,\end{equation}
while the localized subspace is the intersection
\begin{equation}\Hloc = \H_R\cap\H_A.\end{equation}
By ``vanishing quickly,'' we mean that the fields tend to zero fast enough to ensure the existence of advanced and retarded propagators with the properties assumed below. Throughout this section we assume that $\Phi$ satisfies the equations of motion, so the kinetic operator $Q_\Phi$ is nilpotent. We also assume that the action of $Q_\Phi$ is localized (though not necessarily strictly local), so it operates consistently within each subspace $\H_R,\H_A,\Hloc$.   Meanwhile the sigmoid operator maps between the subspaces as 
\begin{equation}\sigma:\H\to \H_R,\ \ \ \ \sigma:\H_A\to \Hloc,\end{equation}
and
\begin{equation}1-\sigma:\H\to \H_A.\ \ \ \ 1-\sigma:\H_R\to\Hloc.\end{equation}
The {\it retarded propagator} is an anticommuting operator at grade $-1$ which acts on the retarded subspace
\begin{equation}\Delta_\Phi^R: \H_R \to \H_R,\end{equation}
and satisfies
\begin{equation}[Q_\Phi,\Delta_\Phi^R] = 1.\label{eq:QD1}\end{equation}
This assumes more structure than in subsection \ref{subsec:SymObs}, where the retarded propagator is only defined on time-localized $Q_\Phi$-invariant states at grade 1. In local field theories with finite speed of propagation, the required extension can be shown to exist. The generalization from localized to retarded states follows a standard argument given in \cite{Bar2} (section 3.3), and for local gauge theories with an
appropriate gauge fixing, the extension to other grades is explained in \cite{Benini}. However, in general we simply assume that a retarded propagator of this kind exists. Because of \eq{QD1} the retarded propagator is a contracting homotopy for $Q_\Phi$, which means that $Q_\Phi$ can have no cohomology in the retarded subspace. Of course we expect cohomology will exist in the entire space $\H$ because there should be nontrivial solutions to the equations of motion. However, these solutions should not exist in $\H_R$ because a physical disturbance should not be able to appear out of the vacuum without cause. In a similar way we define the  {\it advanced propagator} as an anticommuting operator at grade $-1$ which acts on the advanced subspace,
\begin{equation}\Delta_\Phi^A: \H_A \to \H_A,\end{equation}
and satisfies
\begin{equation}[Q_\Phi,\Delta_\Phi^A] = 1.\end{equation}
The reasoning behind the existence of the advanced propagator is analogous to the retarded case. Finally, the {\it causal propagator} is an anticommuting operator at grade $-1$ 
\begin{equation}\Deltacaus_\Phi:\Hloc\to\H,\end{equation} 
given by
\begin{equation}\Deltacaus_\Phi = \Delta^R_\Phi - \Delta^A_\Phi.\end{equation}
The causal propagator acts on the localized subspace, where both advanced and retarded propagators can be simultaneously defined. The output however is generally not localized either in the past or the future. The causal propagator satisfies
\begin{equation}
[Q_\Phi,\Deltacaus_\Phi] = 0.
\end{equation}
It is important to be aware of domains in all of these definitions.

\subsection{Inverse on cohomology}
\label{subsec:iso}

The generating parameter of a symmetry $\xi_\Phi$ is annihilated by $Q_\Phi$, and adding a $Q_\Phi$-exact state only alters the symmetry by a gauge transformation. Therefore $\xi_\Phi$ is naturally interpreted as an element of the cohomology of $Q_\Phi$ at grade 0, $H^0(Q_\Phi)$. The characteristic state of an observable $f_\Phi$ is also annihilated by $Q_\Phi$, and adding a $Q_\Phi$-exact state does not change the value of the observable on-shell \cite{Bernardes2}. Therefore $f_\Phi$ can be interpreted as an element of the cohomology of $Q_\Phi$ at grade $1$. An important nuance, however, is that $f_\Phi$ must be localized in time to ensure that it defines a finite observable. Therefore it must be an element of $\Hloc$, and the relevant cohomology $H^1_\mathrm{loc}(Q_\Phi)$ is computed in $\Hloc$. With this in mind, it is natural to guess that the isomorphism between symmetries and observables is extended to an isomorphism between the cohomology groups $H^0(Q_\Phi)$ and $H^1_\mathrm{loc}(Q_\Phi)$. This is indeed the case. This isomorphism is expressed by a mutually inverse pair of maps 
\begin{equation}H^1_\mathrm{loc}(Q_\Phi)\ \begin{matrix}\xrightarrow{\Deltacaus_\Phi} \\[-8pt] \xleftarrow[{[Q_\Phi,\sigma]}]{} \end{matrix}\  H^0(Q_\Phi),\label{eq:iso}\end{equation}
satisfying 
\begin{subequations}
\begin{align}
[Q_\Phi,\sigma]\Deltacaus_\Phi & =1\ \ \ \mathrm{on}\ H^1_\mathrm{loc}(Q_\Phi),\label{eq:isoA}\\
\Deltacaus_\Phi [Q_\Phi,\sigma ] & = 1\ \ \ \mathrm{on}\ H^0(Q_\Phi),\label{eq:isoB}
\end{align}
\end{subequations}
where $[Q_\Phi,\sigma]$ is the operator that appears in the symplectic structure. Note that these equations imply a little bit more than the existence of an isomorphism between symmetries and observables. This is because an element of $H^0(Q_\Phi)$ defines a tangent vector to phase space, but not all tangent vectors are symmetries of the symplectic structure. Relatedly, not all elements of $H^1_\mathrm{loc}(Q_\Phi)$ define observables because the associated characteristic operators are not cyclic. Therefore the isomorphism between symmetries and observables is a restriction of \eq{iso} which accounts for appropriate cyclicity conditions on the cohomology.  

We start by proving \eq{isoA}. This amounts to a re-expression of the argument given in subsection \ref{subsec:SymObs}. Acting on $Q_\Phi$-invariant states in $\Hloc$ we can compute
\begin{align}
[Q_\Phi,\sigma]\Deltacaus_\Phi & = Q_\Phi\sigma\Deltacaus_\Phi\nonumber\\
& = Q_\Phi \Big(\Delta_\Phi^R - (1-\sigma)\Delta_\Phi^R -\sigma \Delta_\Phi^A\Big)\nonumber\\
& = 1- Q_\Phi \Big((1-\sigma)\Delta_\Phi^R +\sigma \Delta_\Phi^A\Big).\label{eq:isoAcomp}
\end{align}
The output of both terms $(1-\sigma)\Delta_\Phi^R$ and $\sigma \Delta_\Phi^A$ is defined in $\Hloc$. Therefore the second term is trivial in the cohomology $H^1_\mathrm{loc}(Q_\Phi)$, which establishes the result. Next we prove \eq{isoB}. Acting on $Q_\Phi$ invariant states in $\H$ we have
\begin{align}
\Deltacaus_\Phi[Q_\Phi,\sigma] & = \Delta^R_\Phi Q_\Phi \sigma + \Delta^A_\Phi Q_\Phi(1-\sigma) \nonumber\\
& = [\Delta^R_\Phi, Q_\Phi] \sigma + [\Delta^A_\Phi, Q_\Phi](1-\sigma)-Q_\Phi\Big(\Delta^R_\Phi\sigma +\Delta^A_\Phi(1-\sigma) \Big)\nonumber\\
&= 1 - Q_\Phi\Big(\Delta^R_\Phi\sigma +\Delta^A_\Phi(1-\sigma) \Big).\label{eq:isoBcomp}
\end{align}
Because of the insertion of the sigmoid, the retarded and advanced propagators in the second term are evaluated  consistently on their respective domains. Therefore the second term is well-defined and trivial in the cohomology $H^0(Q_\Phi)$, which establishes the result.

\subsection{Peierls complex}
\label{subsec:complex}

The above construction defines an inverse of the causal propagator on cohomology. However, $\Deltacaus_\Phi$ and $[Q_\Phi,\sigma]$ themselves are defined outside the cohomology on more-or-less arbitrary field configurations. Therefore it makes sense to look for a generalization of the isomorphism \eq{iso} that works for arbitrary states in $\H$ without constraints or hidden equivalences. There is a standard construction referenced earlier that works in the absence of gauge symmetry. Here we extend this to an exact sequence that accounts for gauge symmetries and provide an explicit formula for the contracting homotopy. This is what we call the {\it Peierls complex}. 

The Peierls complex involves simultaneously the full vector spaces $\Hloc$ and $\H$ combined as
\begin{equation}\HP = \Hloc\oplus\H.\end{equation}
The grading on $\HP$ is defined in a nontrivial manner. For elements from $\H$ the grade of $\HP$ is identified with that of $\H$. But for elements from $\Hloc$, the grade of $\HP$ is that of $\Hloc$ minus two. The grade $p$ subspace of $\HP$ will then be identified
\begin{equation}\HP^{(p)} =  \Hloc^{(p+2)}\oplus \H^{(p)},\end{equation}
where the superscript indicates the grade in the respective space. The Peierls complex takes the form
\begin{equation}
\cdots \xrightarrow{\ \ \Delta\ \ }\HP^{(-1)}\xrightarrow{\ \ \Delta\ \ }\HP^{(0)}\xrightarrow{\ \ \Delta\ \ }\HP^{(1)} \xrightarrow{\ \ \Delta\ \ }\cdots,
\end{equation}
where $\Delta$ is the differential at grade $1$. Acting on the components of $\HP$ it can be expressed as a matrix
\begin{equation}
\Delta = \left(\begin{matrix}Q_\Phi\  & 0 \\ \Deltacaus_\Phi & Q_\Phi\end{matrix}\right).
\end{equation}
This should primarily be seen as representing the action of the causal propagator. The $Q_\Phi$s appear on the diagonal to ensure that we generate an exact sequence in the end, specifically because the image of $Q_\Phi$ characterizes the kernel of the causal propagator. We can also describe this as a double complex (with two rows) involving $\Hloc$ and $\H$ where $Q_\Phi$ is the horizontal differential and $\Deltacaus_\Phi$ is the differential displayed diagonally in
\begin{equation}
\begin{tikzcd}[column sep=2.5em, row sep=2.5em ]
\cdots \arrow[rd, "\Deltacaus_\Phi"] \arrow[r,"Q_\Phi"]& \Hloc^{(0)} \arrow[rd, "\Deltacaus_\Phi"] \arrow[r,"Q_\Phi"] & \Hloc^{(1)} \arrow[rd, "\Deltacaus_\Phi"] \arrow[r,"Q_\Phi"] & \Hloc^{(2)} \arrow[rd, "\Deltacaus_\Phi"] \arrow[r,"Q_\Phi"] & \cdots \\
\cdots \arrow[r,"Q_\Phi"] & \H^{(-2)} \arrow[r,"Q_\Phi"]  & \H^{(-1)} \arrow[r,"Q_\Phi"] & \H^{(0)} \arrow[r,"Q_\Phi"] & \cdots.
\end{tikzcd}
\end{equation}
This construction is standard in homological algebra (see \cite{Weibel} section 1.5.1), and in that terminology it is known as the \emph{mapping cone} of $\Deltacaus_\Phi$. If there is no gauge symmetry, all vector spaces except those at grade 0 and 1 disappear, and the Peierls complex collapses to 
\begin{equation}
\begin{tikzcd}[column sep=1.8em, row sep=1.8 em ]
0\arrow[r] &\Hloc^{(0)}\arrow[r,"Q_\Phi"] & \Hloc^{(1)} \arrow[rd, "\Deltacaus_\Phi"]  &             & &\\
&             &                                & \H^{(0)} \arrow[r,"Q_\Phi"] & \H^{(1)} \arrow[r] & 0 .
\end{tikzcd}
\end{equation}
This is the standard exact sequence described in \cite{Benini2}. The fact that it is exact implies the isomorphism of cohomologies \eq{iso} above (at least in the absence of gauge symmetry), though the explicit inverse map $[Q_\Phi,\sigma]$ is not yet provided. Originally the inverse map was not known, and proving exactness required carefully checking the kernel and image at each node in the complex. When gauge symmetries are present this becomes a major effort, and is one of the main objectives of the analysis in~\cite{Khavkine} (see section 3.2.3). Other results in \cite{Khavkine} (the {\it splitting formulas} of lemma 2.1) suggested that a proof of exactness may be possible through the construction of a contracting homotopy. Such a construction was later described in the absence of gauge symmetry in \cite{Khavkine2,Lupo}. 

Now we describe the full contracting homotopy. This is an operator $\Omega$ at grade $-1$ which represents the reverse arrows in the  Peierls complex, 
\begin{equation}
\cdots\  \begin{matrix}\xrightarrow{\ \ \Delta\ \ } \\[-8pt] \xleftarrow[\ \ \Omega\ \ ]{} \end{matrix}
\ \HP^{(-1)}\ \begin{matrix}\xrightarrow{\ \ \Delta\ \ } \\[-8pt] \xleftarrow[\ \ \Omega\ \ ]{} \end{matrix}\ \HP^{(0)}\ \begin{matrix}\xrightarrow{\ \ \Delta\ \ } \\[-8pt] \xleftarrow[\ \ \Omega\ \ ]{} \end{matrix}\ \HP^{(1)}\ \begin{matrix}\xrightarrow{\ \ \Delta\ \ } \\[-8pt] \xleftarrow[\ \ \Omega\ \ ]{} \end{matrix}\ \cdots,
\end{equation}
and satisfies 
\begin{equation}[\Delta,\Omega] = 1.\end{equation}
The solution we find is displayed as matrix acting on the components of $\HP$ as
\begin{equation}
\Omega = \left(\begin{matrix} (1-\sigma)\Delta^R_\Phi + \sigma \Delta^A_\Phi & [Q_\Phi,\sigma], \\ (\Delta_\Phi^A)^3 - (\Delta_\Phi^R)^3 & \Delta_\Phi^R \sigma + \Delta_\Phi^A (1-\sigma)\end{matrix}\right).
\end{equation}
The operator in the upper right hand corner comes from the symplectic form and will invert the causal propagator on cohomology. It is important to check that the components of this matrix map correctly between the relevant subspaces 
\begin{align}
(1-\sigma)\Delta^R_\Phi + \sigma \Delta^A_\Phi: & \ \Hloc\to\Hloc, \nonumber\\
[Q_\Phi,\sigma]: & \ \H\to\Hloc,\nonumber\\
(\Delta_\Phi^A)^3 - (\Delta_\Phi^R)^3: & \ \Hloc\to\H, \nonumber\\
\Delta_\Phi^R \sigma + \Delta_\Phi^A (1-\sigma):& \ \H \to \H.
\end{align}
If we label the matrix components of $\Omega$ as
\begin{equation}
\Omega = \left(\begin{matrix}\Omega_{11} &\Omega_{12} \\ \Omega_{21} & \Omega_{22}\end{matrix}\right),
\end{equation}
the condition $[\Delta, \Omega]=1$ expands into four equations:
\begin{subequations}
\begin{align}
[Q_\Phi,\Omega_{12}] & = 0,\\
\Omega_{12}\Deltacaus_\Phi  & = 1-[Q_\Phi,\Omega_{11}],\\
\Deltacaus_\Phi \Omega_{12} & = 1-[Q_\Phi, \Omega_{22}],\\
\Deltacaus_\Phi \Omega_{11} & +[Q_\Phi,\Omega_{21}]+ \Omega_{22}\Deltacaus_\Phi=0.\label{eq:527d}
\end{align}
\end{subequations}
The first equation implies that $[Q_\Phi,\sigma]$ is a well defined operator on cohomology. The second two equations are actually the result of the calculations \eq{isoAcomp} and \eq{isoBcomp} which demonstrate that $\Deltacaus_\Phi$ and $[Q_\Phi,\sigma]$ are inverses on cohomology.  This follows from a standard result from homological algebra, which states that the cone complex of a map has a contracting homotopy if and only if the map has a two-sided inverse up to homotopy, see theorem 4.2.10 of \cite{Spanier}. 

The final equation fits nicely with homological algebra but its interpretation is perhaps of secondary importance. However, it is the most nontrivial identity as it involves the rather peculiar difference between cubes of advanced and retarded propagators. For the sake of interest it is worth checking explicitly. Plugging in the components of $\Omega$ to the left hand side of \eq{527d},
\begin{equation}
\Deltacaus_\Phi\Big((1-\sigma)\Delta^R_\Phi + \sigma \Delta^A_\Phi\Big)+[Q_\Phi,(\Delta_\Phi^A)^3 - (\Delta_\Phi^R)^3]+\Big(\Delta_\Phi^R \sigma + \Delta_\Phi^A (1-\sigma)\Big)\Deltacaus_\Phi.\label{eq:5.26}
\end{equation}
Because $Q_\Phi$ and $\Delta_\Phi^A$ anticommute,
\begin{align}
[Q_\Phi,(\Delta_\Phi^A)^3] & = [Q_\Phi,\Delta_\Phi^A](\Delta_\Phi^A)^2-\Delta_\Phi^A[Q_\Phi,\Delta_\Phi^A]\Delta_\Phi^A+(\Delta_\Phi^A)^2[Q_\Phi,\Delta_\Phi^A]\nonumber\\
& = (\Delta_\Phi^A)^2-(\Delta_\Phi^A)^2+(\Delta_\Phi^A)^2\nonumber\\
&= (\Delta_\Phi^A)^2,
\end{align}
and the difference of cubes in the second term becomes a difference of squares. For the first term and second terms we can rewrite
\begin{subequations}
\begin{align}
\Deltacaus_\Phi\Big((1-\sigma)\Delta^R_\Phi + \sigma \Delta^A_\Phi\Big) & = \Deltacaus_\Phi\Big(\Delta^R_\Phi -\sigma \Deltacaus_\Phi\Big) \nonumber\\
& = (\Delta^R_\Phi)^2 -\Delta^A_\Phi\Delta^R_\Phi - \Deltacaus_\Phi\sigma \Deltacaus_\Phi,\\
\Big(\Delta_\Phi^R \sigma + \Delta_\Phi^A (1-\sigma)\Big)\Deltacaus_\Phi & = \Big(\Delta_\Phi^A+\Deltacaus_\Phi \sigma \Big)\Deltacaus_\Phi\nonumber\\
& = -(\Delta_\Phi^A)^2+\Delta_\Phi^A\Delta_\Phi^R+\Deltacaus_\Phi \sigma \Deltacaus_\Phi.
\end{align}
\end{subequations}
Putting everything together 
\begin{align}
\Big((\Delta^R_\Phi)^2 -\Delta^A_\Phi\Delta^R_\Phi - \Deltacaus_\Phi\sigma \Deltacaus_\Phi\Big) + \Big((\Delta_\Phi^A)^2-(\Delta_\Phi^R)^2\Big) + \Big(-(\Delta_\Phi^A)^2+\Delta_\Phi^A\Delta_\Phi^R+\Deltacaus_\Phi \sigma \Deltacaus_\Phi\Big).
\end{align}
One can inspect that all of the terms cancel consistent with \eq{527d}.

Though it is fairly clear already from the calculations in subsection \ref{subsec:iso}, it is interesting to note that part of the contracting homotopy identity implies that the isomorphism
\begin{equation}
        H^{p+1}_\mathrm{loc}(Q_\Phi)\
\begin{matrix}\xrightarrow{\Deltacaus_\Phi} \\[-8pt]
\xleftarrow[{[Q_\Phi,\sigma]}]{} \end{matrix}\  H^p(Q_\Phi)
\label{eq:iso-all}
\end{equation}
holds in all grades, not just $p=0$. The cohomologies $H^p(Q_\Phi)$ in
$p<0$ grades can be related to (higher) rigid symmetries of the theory
and its (lower form degree) conserved currents, while those in $p>0$
grades can be related to quantization anomalies \cite{Barnich2}. The above
isomorphism appears to be interesting and does not seem to have been much
noticed in the literature. Its consequences are yet to be explored.

Considering the amount of work that has been invested in developing an exact sequence of this kind, the simplicity of the Peierls complex is remarkable. One can ask why it was not recognized before. Perhaps one reason is that the symplectic form is not obviously realized as an operator on~$\H$. In the standard description the symplectic form is displayed as an integral of a current density over an initial value surface. If compelled this could be understood as an operator on $\H$ characterized by a delta function singularity that localizes to the initial value surface. This however requires working with a singular space of fields. The sigmoid softens the time slice and makes it more natural to think about the symplectic structure as an operator on $\H$. Something related to the Peierls complex appears in~\cite{Benini}, but it is embedded in a highly technical formalism.

\subsection*{Acknowledgments}

AHF thanks Mukund Rangamani for conversations. TE and AHF thank David Gross for hospitality at the KITP while carrying out part of this work. The work of VB and TE was supported by the European Structural and Investment Funds and the Czech Ministry of Education, Youth and Sports (project No. FORTE—CZ.02.01.01/00/22\_008/0004632). The work of AHF is supported by the U.S. Department of Energy, Office of Science, Office of High Energy Physics of U.S. Department of Energy under grant Contract Number DE-SC0009999, and the funds from the University of California.  IK was supported by the Czech Science Foundation (project GA25-15544S) and the Czech Academy of Sciences (Research Plan RVO: 67985840). This research was supported in part by grant NSF PHY-2309135 and the Gordon and Betty Moore Foundation Grant No. 2919.02 to the Kavli Institute for Theoretical Physics (KITP).

\end{document}